\documentclass[12pt,preprint]{aastex}
\usepackage{amsmath,amssymb,graphicx,natbib}

\renewcommand{\vec}[1]{\mathbf{#1}}

\newcommand{\ny}{\hat{\mathbf{e}}_y}
\newcommand{\nz}{\hat{\mathbf{e}}_z}

\newsavebox{\astrutbox}
\sbox{\astrutbox}{\rule[-5pt]{0pt}{20pt}}

\newcommand{\vv}{{\mathbf v}}

\newcommand{\BB}{{\mathbf B}}

\newcommand{\ez}{{\mathbf e_z}}

\newcommand{\OO}{{\mathbf \Omega}}

\newcommand{\gapprox}{\lower.4ex\hbox{$\;\buildrel >\over{\scriptstyle\sim}\;$}}
\newcommand{\lapprox}{\lower.4ex\hbox{$\;\buildrel <\over{\scriptstyle\sim}\;$}}

\bibpunct[; ] {(}{)}{,}{}{}{,}
\slugcomment{To be submitted to ApJ.} 
\shorttitle{Magetorotational turbulence in stratified boxes} 
\shortauthors{Bodo, Cattaneo, Mignone \& Rossi} 

\begin{document} 
 
\title{Magnetorotational turbulence in stratified shearing boxes with perfect gas equation of state and finite thermal diffusivity} 
 
 \author{G. Bodo\altaffilmark{1},        
         F. Cattaneo\altaffilmark{2}, 
         A. Mignone\altaffilmark{3},
         P. Rossi\altaffilmark{1} }
  
 \altaffiltext{1}{INAF, Osservatorio Astronomico di Torino, Strada Osservatorio 20, Pino Torinese, Italy}
 
 \altaffiltext{2}{The Computation Institute, The University of Chicago, 
              5735 S. Ellis ave., Chicago IL 60637, USA}

 \altaffiltext{3}{Dipartimento di Fisica Generale, Univesita di Torino, via Pietro Giuria 1, 10125 Torino, Italy} 
 
\begin{abstract} 

We present a numerical study of turbulence and dynamo action in stratified shearing boxes with zero mean magnetic flux. We assume that the fluid obeys the perfect gas law and has finite (constant) thermal diffusivity. The calculations begin from an isothermal state spanning three scale heights above and below the mid-plane. After a long transient the layers settle to a stationary state in which thermal losses out of the boundaries are balanced by dissipative heating. We identify two regimes.  A conductive regime in which the heat is transported mostly by conduction  and the density decreases with height. In the limit of large thermal diffusivity this regime resembles the more familiar isothermal case.  Another--the convective regime-- observed at smaller values of the thermal diffusivity, in which the layer becomes unstable to overturning motions, the heat is carried mostly by advection and the density becomes nearly constant throughout the layer. In this latter constant-density regime we observe evidence for large-scale dynamo action leading to a substantial increase in transport efficiency relative to the conductive cases. 

\end{abstract} 

\section{Introduction}
The origin of turbulence and enhanced angular momentum transport in accretion flows is a fascinating problem of considerable importance in astrophysics. It is commonly believed that the magneto-rotational instability (MRI), in one form or another,  plays a fundamental role in destabilizing the basic quasi-Keplerian flow.  When a net magnetic flux is present the MRI sets in as a classical linear instability with a well defined growth rate and characteristic wavenumber \citep{Balbus91}; the turbulence then develops from the nonlinear evolution of this instability. When there is no net flux the problem is more complicated and the turbulence must develop from a nonlinear subcritical instability. In this case, the problem becomes fundamentally one  of establishing what form of dynamo action can be sustained in a disc. Much of what is currently known about dynamo action in accretion flows is based on  numerical studies formulated within the framework of the shearing-box approximation \citep{Hawley95}. The simplest set up, both conceptually and numerically, consists of an unstratified, isothermal shearing-box with periodic boundary conditions in the vertical direction. It is now well established that this configuration suffers from the so-called convergence problem.  As the magnetic diffusivity decreases, or equivalently the resolution increases, the Maxwell stresses decrease, eventually to become negligible  \citep[][see however Fromang 2010 for a different view]{Fromang07, Pessah07, Guan09, Simon09, Bodo11}. The cause for this  ``non convergence" has been attributed to the the lack of a characteristic outer scale in the periodic, unstratified problem \citep[for a discussion see][]{Bodo11}. The next step towards more realistic simulations is to retain the shearing-box geometry but with the inclusion  of vertical gravity, and consequently, stratification. This introduces a characteristic length--the scale height--that may help to remedy the convergence problem \citep{Davis10, Shi10, Oishi11}. Whether this is the case or not, at the moment, remains an open question. Certainly, in the stratified cases the solutions manifest a richness both in space and time  that is absent in the unstratified cases \citep{Gressel10, Guan11, Simon12}. It is important to note that most of these studies adopt an isothermal equation of state; the resulting density distribution is correspondingly close to a Gaussian with most of the mass concentrated near the mid-plane, and tenuous, low density regions above and below. This leads to very different dynamo processes operating in the mid-plane and in the overlying regions.  Although an isothermal formulation is conceptually simple and easy to implement numerically, it neglects the possibly important process of turbulent heating by viscous and Ohmic dissipation. It can be argued that in an optically thin environment turbulent heating may not be important since the energy can easily escape without substantially heating the ambient plasma. However, this is definitely not the case in an optically thick environment. In this case the plasma will be heated locally and the final thermal structure will be determined by a balance between energy deposition and energy transport. In this case, it is possible that substantial departures may develop from the isothermal case that, in turn may impact the operation of the dynamo. 
Some of these issues have been addressed  by Hirose and collaborators  \citep{ Hirose06, Hirose09, Blaes11},  who have considered radiation dominated discs and have included a sophisticated treatment of the radiation field, and also by the works of Flaig and collaborators \citep{Flaig10, Flaig12} whose models of proto-planetary discs include partial ionization, chemical networks and heat transport in the radiative conduction approximation. All these works indicate that turbulent heating can indeed be important. Here, we also address the problem of turbulent heating but in the somewhat simpler case of a fully ionized, pressure dominated disc. Our intention is to provide a bridge between the works of Hirose et al. and Flaig et al. and those based on the isothermal equation of state. To this end we consider a stratified shearing-box with a perfect gas equation of state and finite (constant) thermal diffusivity. The objective is to study how the basic state and the corresponding dynamo action changes as the thermal diffusivity is varied. In this work we  deliberately keep the formulation as simple as possible in order to highlight some of the  basic underlying physical processes. 

\section{Formulation} \label{formulation} 
Our objective is to provide a simple model in which the effects of dissipative heating can be studied. In particular we want to assess how these processes, together with thermal transport lead to departures from the more familiar isothermal cases. We assume that the plasma is optically thick and approximate the radiative transport by a diffusion process which we model by a thermal conduction term in the energy equation. In the spirit of keeping things as simple as possible, and in order to capture  more easily the general properties of the solutions, we make  further simplifications, by neglecting the dependencies on density and temperature resulting from the diffusion approximation to the radiative transport equation, and assuming a constant thermal diffusivity. A more realistic treatment of the radiation will be considered in future work.

We perform three-dimensional, numerical simulations of a perfect gas with thermal conduction in a shearing box with vertical gravity.  A detailed presentation of the shearing box approximation can be found in \citet{Hawley95}. The Magneto-Hydro-Dynamics (MHD) shearing-box equations, including  vertical gravity  and thermal conduction can be written as:

\begin{equation}
\frac{\partial \rho}{\partial t} + \nabla \cdot \left( \rho \vv \right) = 0,
\label{eq:mass}
\end{equation}

\begin{equation}
\frac{\partial \vv}{\partial t} + \vv \cdot \nabla \vv + 2 \Omega \times \vv = \frac{\BB \cdot \nabla \BB}{4 \pi \rho} - \frac{1}{\rho} \nabla 
\left(  \frac{\BB^2}{8 \pi} + P \right) - \nabla \left( 2 A \Omega x^2  + \frac{1}{2} \Omega^2 z^2 \right),
\label{eq:momentum}
\end{equation}

\begin{equation}
\frac{\partial \BB}{\partial t} - \nabla \times \left( \vv \times \BB \right) = 0,
\label{eq:induction}
\end{equation}

\begin{equation}
\frac{\partial E}{\partial t} + \nabla \cdot [(E + P_T) \vv + (\vv \cdot \BB) \BB - k \nabla T] = 0 ,
\label{eq:energy}
\end{equation}
where $\BB$, $\vv$, $\rho$ and $P$ denote, respectively,   the magnetic  field intensity, the velocity, the density and the thermal pressure; $E$ is the total energy density, $P_T$ is the total (thermal plus magnetic) pressure and $k$ is the thermal conductivity. The local angular velocity $\OO = \Omega \ez$ and the shear rate 
 \begin{equation}
A \equiv \frac{R}{2} \frac{\partial \Omega}{\partial R}
\end{equation}
are assumed constant.  For a Keplerian disk  $A = -(3/4) \Omega$.   The system is closed by the equation of state for a perfect gas:  
\begin{equation}
P = \rho T
\end{equation}
where we have absorbed the perfect gas constant in the definition of the temperature.
The thermal conductivity can be written as
\begin{equation}
k = \frac{5}{2}  \kappa \rho
\end{equation}
where $\kappa$ is the thermal diffusivity, which, as discussed above, we assume  to be constant, and the factor of $5/2$ is appropriate for a gas with three degrees of freedom.

We start our simulations from a state with a uniform shear flow,  $\vec{v} = -2 A x\ny$,  and  density and pressure distributions that satisfy  vertical hydrostatic balance with constant temperature  $T_0$. With these conditions the initial density has a Gaussian profile given by
\begin{equation}
\rho = \rho_0 \exp(- \Omega ^2 z^2 / 2 T_0),  \
\end{equation}
where $\rho_0$ is the value of density on the equatorial plane.

If the MRI develops to substantial amplitude, this initial state will be driven away from thermal equilibrium by the energy input from dissipative processes. The temperature in the equatorial regions  will  progressively increase and a thermal gradient will be established until a new equilibrium is reached whereby the energy input is  balanced by thermal losses at the upper and lower boundaries. As we shall see, the new equilibrium can be quite different from the initial isothermal state and is determined self-consistently by the heating associated with the process of angular momentum transport by the MRI. We note here that, in our current formulation, we do not include viscous and Ohmic dissipation explicitly. The heating of the fluid occurs because of numerical dissipation together with a conservative formulation of the total energy equation. The latter requires that whatever kinetic or magnetic energy is lost by dissipative processes  it be re-introduced in the form of internal energy (heating). 

The computational domain covers the region $L_x \times L_y \times L_z$, where $L_x = H$, $L_y = \pi H$ and $L_z = 6H$,
where
\begin{equation}
H = \frac{\sqrt{2 T_0}}{\Omega}
\end{equation}
is the pressure scale height in the initial isothermal state. In the vertical direction the box is symmetric with respect to the equatorial plane $z = 0$,  where gravity changes sign. Numerically, the domain is covered by a grid of $32 \times 96 \times 192$ grid points.  
A  magnetic field of the form
\begin{equation}
\BB = B_0 \sin \left( \frac{2 \pi x}{H} \right) \nz
\end{equation}
is imposed initially
where $B_0$ corresponds to the ratio between thermal and magnetic pressure and has a value of $1600$. Clearly, there is no net magnetic flux threading the box.  In addition we introduce random noise in the $y$ component of the velocity in order to destabilize the system. 

Following common practice, we assume periodic  boundary conditions in the $y$ direction and shear periodic conditions in the $x$ direction.  In the vertical direction,  we assume that  the upper and lower boundaries ($z = \pm 3H$), are impenetrable and stress free, giving $v_z = 0$ , $\partial v_x / \partial z = \partial v_y / \partial z = 0$, and also that the magnetic field is purely vertical, giving  $\partial B_z / \partial z = 0$, $B_x = B_y = 0$. We should note that these conditions allow a net flux of magnetic helicity through the boundaries with, possibly, important consequences  to the dynamo processes  \citep{VC01, Kapyla10}.
Finally, we assume that the boundaries are in hydrostatic balance, and that the temperature is constant and equal to $T_0$; thus
\begin{equation}
\frac {\partial p_T}{\partial z} = \mp  3 \rho \Omega^2 H, \qquad \qquad T=T_0.
\end{equation}
All  simulations are carried out with the PLUTO code \citep{Mignone07}, with a second order accurate scheme, HLLD Riemann solver and an explicit treatment of thermal conduction.

\section{Results} \label{results} 
\begin{figure}[htbp]
   \centering
   \includegraphics[width=10cm]{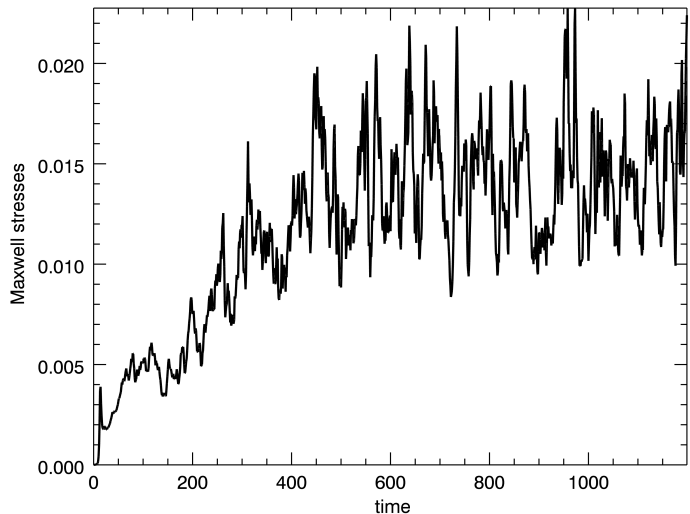} 
   \caption{Time history of the Maxwell stresses averaged over the computational box for the case $\kappa = 2 \times 10^{-2}$. }
   \label{fig:maxwtime}
\end{figure}

\begin{figure}[htbp]
   \centering
   \includegraphics[width=10cm]{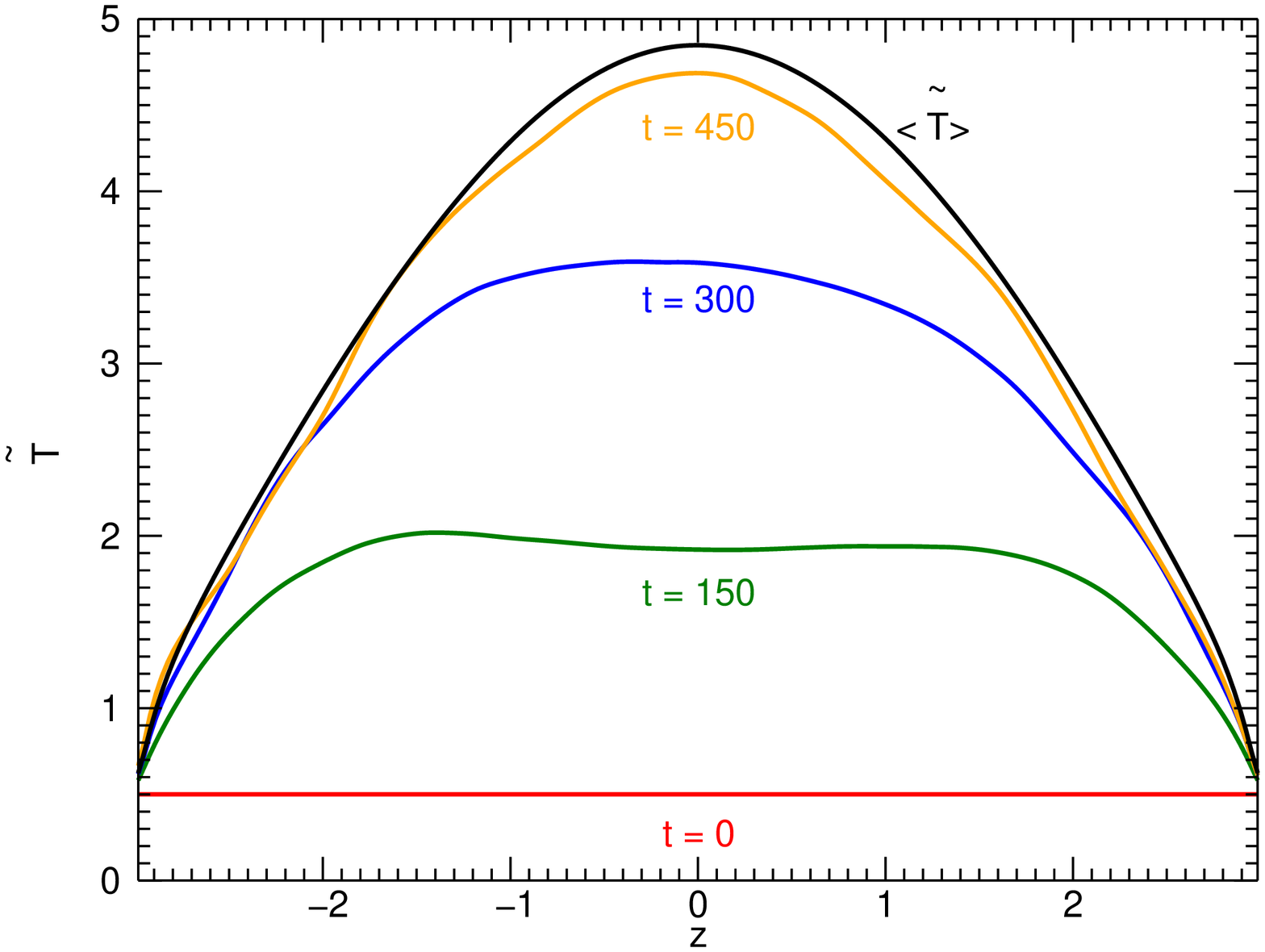} 
   \caption{Temperature averaged over horizontal planes, $\tilde T$ as a function of the vertical coordinate $z$ for the case $\kappa = 2 \times 10^{-2}$. The different curves refer to different times, as indicated by the labels, and for comparison we plot also  the time averaged distribution $ \langle \tilde T \rangle$  in the steady state. }
   \label{fig:avt_time}
\end{figure}

\begin{figure}[htbp]
   \centering
   \includegraphics[width=10cm]{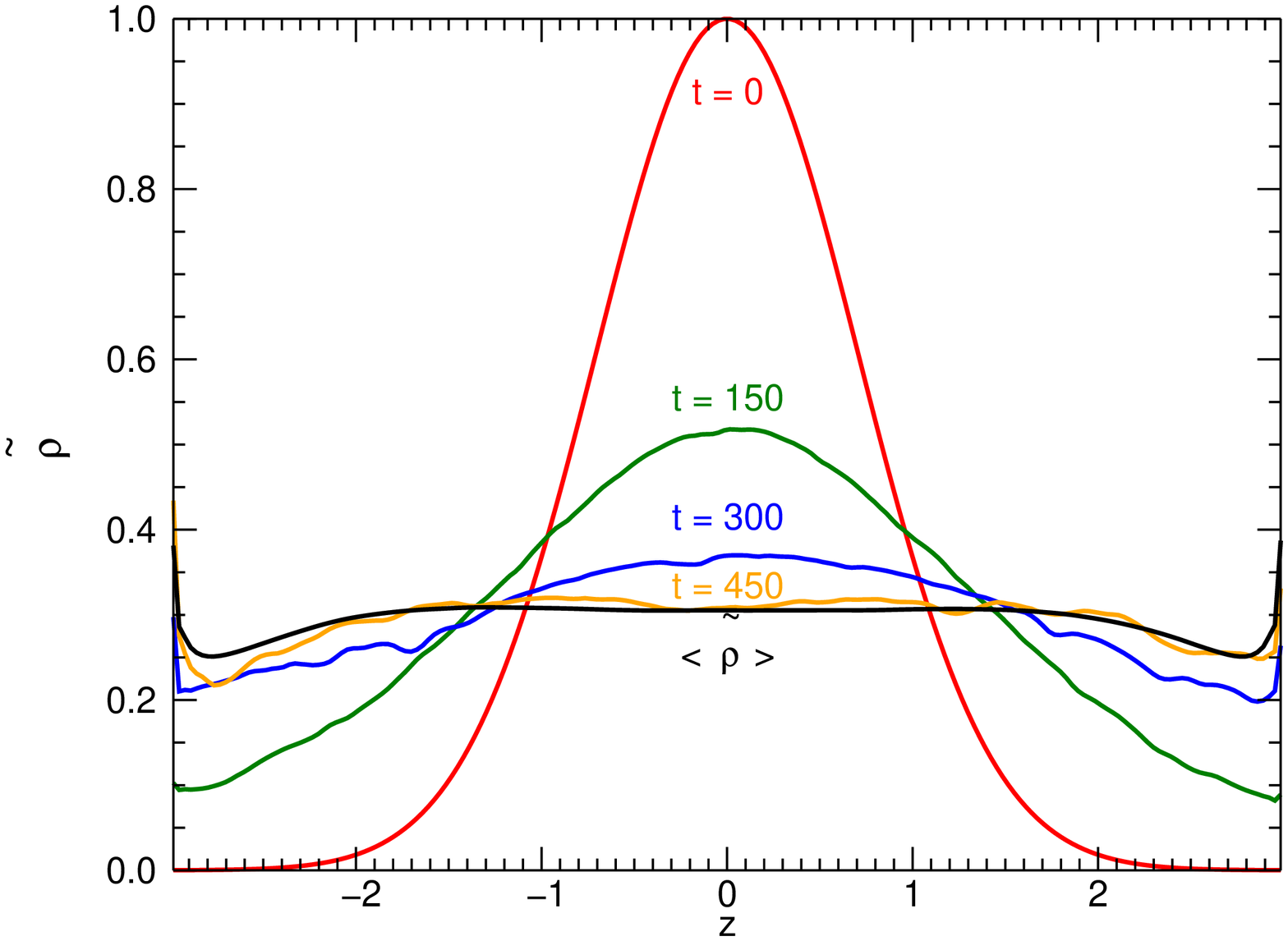} 
   \caption{Density averaged over horizontal planes, $\tilde \rho$ as a function of the vertical coordinate $z$ for the case $\kappa = 2 \times 10^{-2}$. The different curves refer to different times, as indicated by the labels, and for comparison we plot also the time averaged distribution $\langle \tilde \rho \rangle$  in the  steady state. }
   \label{fig:avrho_time}
\end{figure}

We now describe the development of MRI driven turbulence from an initially isothermal state. Hereinafter, and unless otherwise specified, when presenting the numerical results, we adopt $\Omega^{-1}$ as the unit of time, $H$ as the unit of length, and the mid-plane density in the initial isothermal state $\rho_0$, as the unit of density and,  since  $H$ is  our unit of length, we have $T_0=1/2$. 
Following the initial perturbations, a sub-critical instability sets in leading to the generation of magnetic fields and the development of turbulence. Dissipative processes heat the plasma driving the system away from the initial isothermal state. Eventually the system reaches a stationary state in which the heat generated by the turbulence is balanced by the heat lost through the upper and lower boundaries. 
Locally, the balance is between the volumetric heat production and the divergence of the heat flux that can arise both by thermal conduction and turbulent transport.  The relative importance of these two processes depends on the value of the thermal diffusivity, which, here, is expressed in units of the product of the scale-height and the isothermal sound speed, i.e. it has the form of  
an inverse P\'eclet number. A typical evolution for a case with $\kappa = 2 \times 10^{-2}$ can be followed in Fig. \ref{fig:maxwtime} where we show the time history of the Maxwell stresses averaged over the entire computational domain. Clearly, there is a long adjustment phase lasting approximately 500 time-units after which the system settles into a stationary state in which the stresses remain strongly fluctuating but with a well defined (time) average value. The corresponding thermal history can be assessed by inspection of Figs. \ref{fig:avt_time} and \ref{fig:avrho_time}, showing respectively, the horizontally averaged temperature, $\tilde T(z)$ and density $\tilde \rho(z)$, at several times. The asymptotic profiles in the stationary state (obtained by time averaging from $t = 500$ to the end of the simulation, $t = 2000$) are denoted by angle brackets. Clearly the increase in the Maxwell stresses are accompanied by the  heating of the central regions leading to the establishment of a nearly parabolic profile in temperature. We note a corresponding dramatic change in the density distribution that evolves from the initial Gaussian profile to an almost constant distribution at later times. 

The development of a constant density state is somewhat remarkable, and deserves further investigation. At first sight it may appear as the result of a fortuitous choice of thermal diffusivity. As we shall see presently, this is not entirely the case. 
In the stationary state the average temperature and density are related by the condition of hydrostatic balance, which we write here in dimensional form, and for simplicity we only consider  $z > 0$;
\begin{equation}
\frac{1}{\rho} \frac{d \rho}{d z} = \frac{1}{T} \left( -\Omega^2 z - \frac{d T}{d z} \right) .
\label{eq:hydeq}
\end{equation}
Clearly, whether the density decreases upwards, increases upwards, or remains constant depends on the relative magnitude of the two terms in the brackets on the RHS of (\ref{eq:hydeq}).  The first term is a fixed linear function of $z$. The second--the temperature gradient--is negative since the layer is heated from within, but its magnitude depends on a balance between local heat production rate and local heat transport. 
To a first approximation, one could assume that the energy production rate should be independent of the thermal diffusivity $\kappa$. This is not unreasonable, since the production rate is driven by turbulent dissipation, which in turn depends solely on the efficiency of the MRI. This being the case, the magnitude of the temperature gradient could be made arbitrarily small by choosing a large value of $\kappa$. Clearly, if the thermal diffusivity is huge, thermal conduction can easily transport all the generated heat along  very shallow gradients. The temperature will be nearly constant, the density will rapidly decrease upward in accordance to (\ref{eq:hydeq}), and resemble the isothermal distribution. By contrast, if $\kappa$ is tiny the temperature gradients required to carry the heat will be huge (in absolute value), the RHS of (\ref{eq:hydeq}) will be positive and the density will rapidly increase with height. However, this configuration with a density inversion is strongly unstable to Rayleigh-Taylor type instabilities. The resulting overturning motions will both carry the heat more efficiently than thermal conduction, and homogenize the mass towards a constant density state. Thus we can conjecture the existence of a critical value of $\kappa=\kappa_{crit}$ above which the transport is mostly conductive, the layers have a density decreasing with height and a stratification approaching that of an isothermal layer  in the limit of large $\kappa$ (conductive states). For  $\kappa \ll \kappa_{crit}$, the heat transport is mostly advective, and the density is approximately constant (convective states).

\begin{figure}[htbp]
   \centering
   \includegraphics[width=10cm]{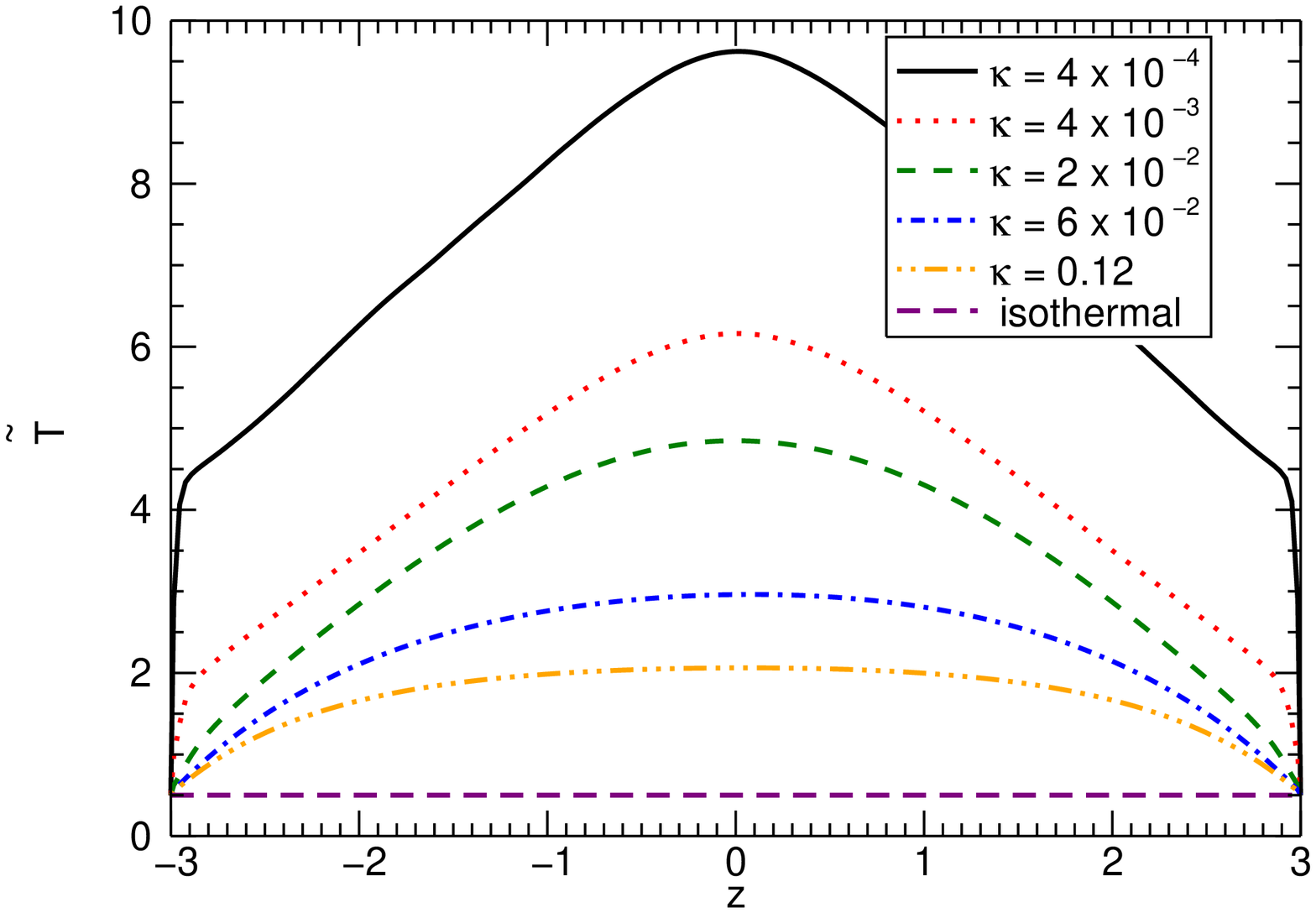} 
   \caption{Plot of $ \langle \tilde T \rangle $ as a function of $z$. The different curves refer to different values of $\kappa$,  as shown in the legend. For comparison we plot also the isothermal case}
   \label{fig:avt_z}
\end{figure}

\begin{figure}[htbp]
   \centering
   \includegraphics[width=10cm]{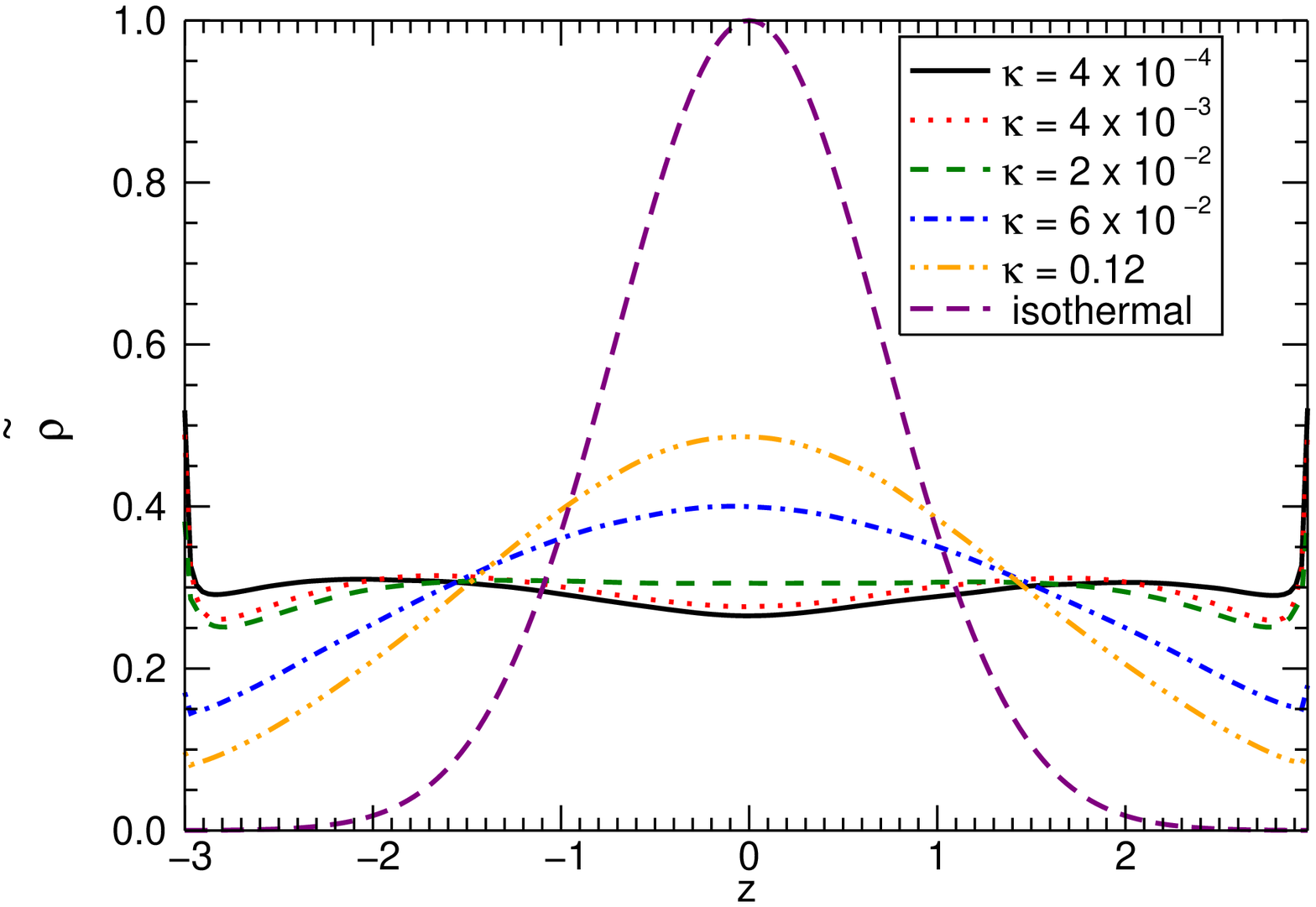} 
   \caption{Plot of $ \langle \tilde \rho \rangle  $ as a function of $z$. The different curves refer to different values of $\kappa$,  as shown in the legend. For comparison we plot also the isothermal case}
   \label{fig:avrho_z}
\end{figure}

Some of these ideas can be easily verified by considering a series of calculations with varying thermal diffusivity. The results are summarized in Figs. \ref{fig:avt_z} and \ref{fig:avrho_z} where we show the steady state horizontally averaged temperature and density distributions for  different values of $\kappa$. As expected, the temperature gradient is always negative ($z>0$), its magnitude increases with decreasing $\kappa$ as does the overall temperature of the layer. For large values of $\kappa$ the temperature distributions have an approximate parabolic profile and the density decreases upwards. 

\begin{figure}[htbp]
   \centering
   \includegraphics[width=10cm]{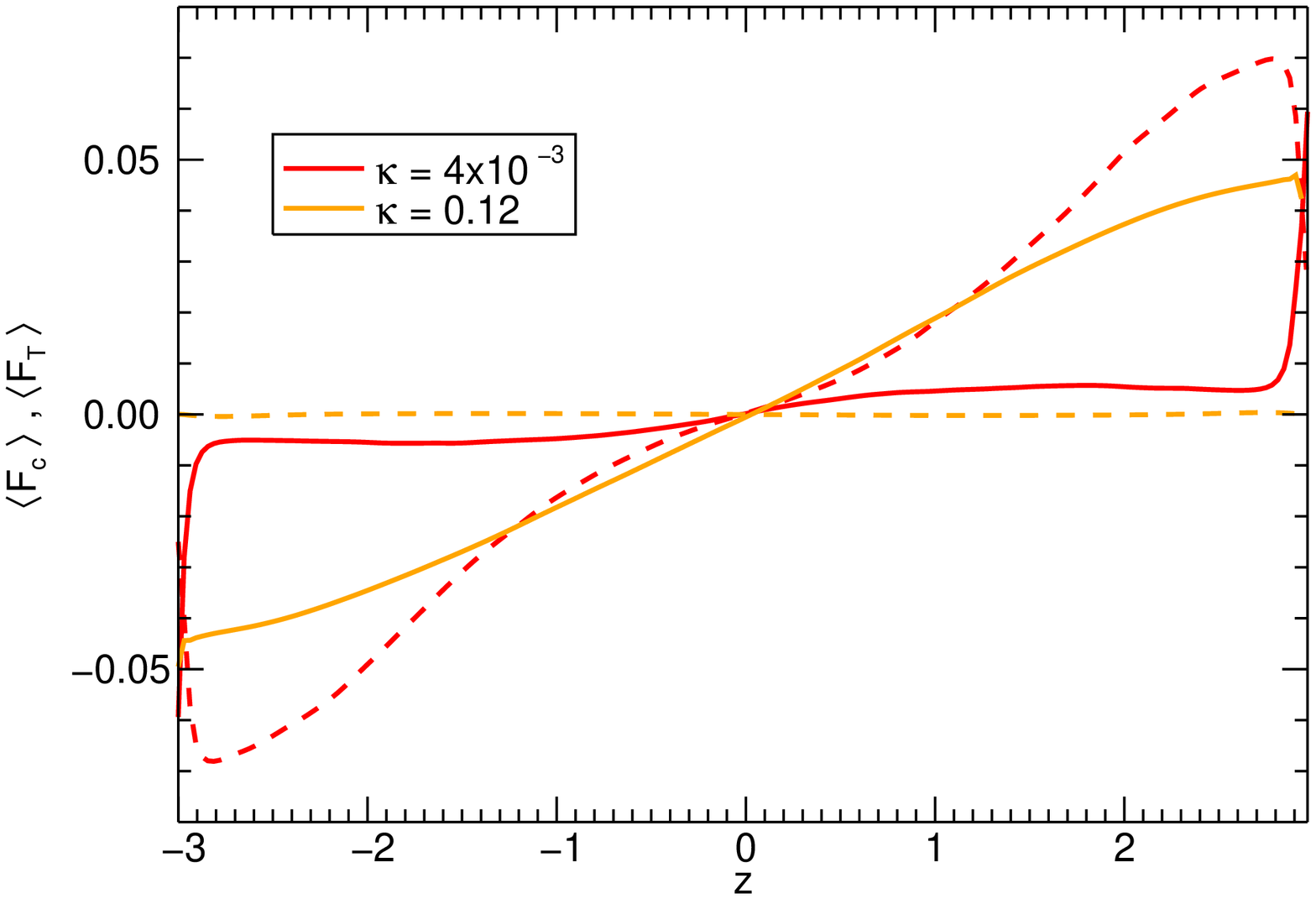} 
   \caption{Plot of $ \langle F_c \rangle $ and $ \langle F_T \rangle$ as functions of $z$. The solid curves show $\langle F_c \rangle $ while the dashed curves show $ \langle F_T \rangle $, the different colors refer to different values of $\kappa$ as indicated in the legend.  }
   \label{fig:th_fluxes}
\end{figure}

For small values of $\kappa$ the temperature in the interior approaches a ``tent"  profile with a parabolic shape near the equator then a linear decrease over most of the domain and thin boundary layers at the edges. As $\kappa$ decreases the profiles move up retaining their shape but producing progressively thinner boundary layers. 
The corresponding density profiles confirm the establishment of a constant density state  that becomes asymptotically independent of $\kappa$. From Fig. \ref{fig:avrho_z}, we can estimate that the critical value of $\kappa$ for a transition from conductive to convective regimes, in this setup,  satisfies
\begin{equation}
\kappa_{crit} \approx 2 \times 10^{-2}.
\label{eq:kappa-crit}
\end{equation}
Our conjecture that as $\kappa$ crosses its critical value the vertical heat transport changes from conductively dominated to advectively dominated can also be verified by considering the horizontally averaged conductive, and advective fluxes that can be defined, respectively, as
\begin{equation}
F_c = - \frac{5}{2} \kappa \rho \frac{d \tilde T}{d z} ,
\end{equation}
and 
\begin{equation}
F_T = \frac{1}{L_x L_y} \int \frac{5}{2}  \rho v_z (T - \tilde T) dx dy .
\end{equation}
Their values in the stationary state for the two extreme values of $\kappa$ are shown in Fig. \ref{fig:th_fluxes}. The roles of the two types of flux practically reverse. For $\kappa=0.12$ the transport is entirely conductive and advection is negligible; for $\kappa=4\times 10^{-3}$ heat conduction is negligible, except in the boundary layers, and  all of the flux is carried by advection. It is interesting to note that near the equator where the advective flux is small--it is actually zero at the equator--the density displays a weak inversion. This is related to the absence of gravity near the equator to drive Rayleigh-Taylor instabilities.

\begin{figure}[htbp]
   \centering
   \includegraphics[width=10cm]{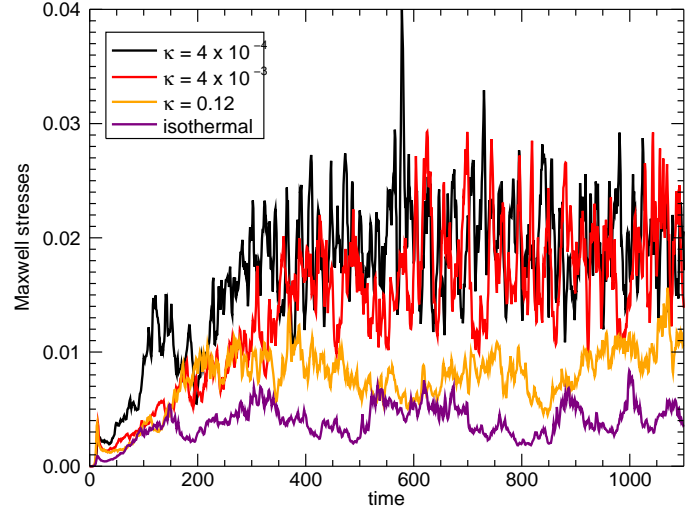} 
   \caption{Time histories of the volume averaged Maxwell stresses for different values of the thermal diffusivity $\kappa$. The values of $\kappa$ are shown in the legend. }
   \label{fig:maxw}
\end{figure}

The existence of two regimes, conductive and convective,  with strikingly different vertical structures is likely to lead to correspondingly different dynamo actions. A measure of these differences can be assessed by inspection of Fig. \ref{fig:maxw} where the domain averaged Maxwell stresses are shown as a function of time for different values of $\kappa$. The corresponding curve for an isothermal case is also included for comparison. Clearly, the angular momentum transport efficiency increases with decreasing $\kappa$ and eventually saturates in the convective regime. It is natural to assume that once the heat transport is mostly advective further decreases in thermal diffusivity will not make any difference. What is remarkable is the difference between the convective cases and the purely isothermal one, with the latter being strikingly smaller. 

\begin{figure}[htbp]
   \centering
   \includegraphics[width=10cm]{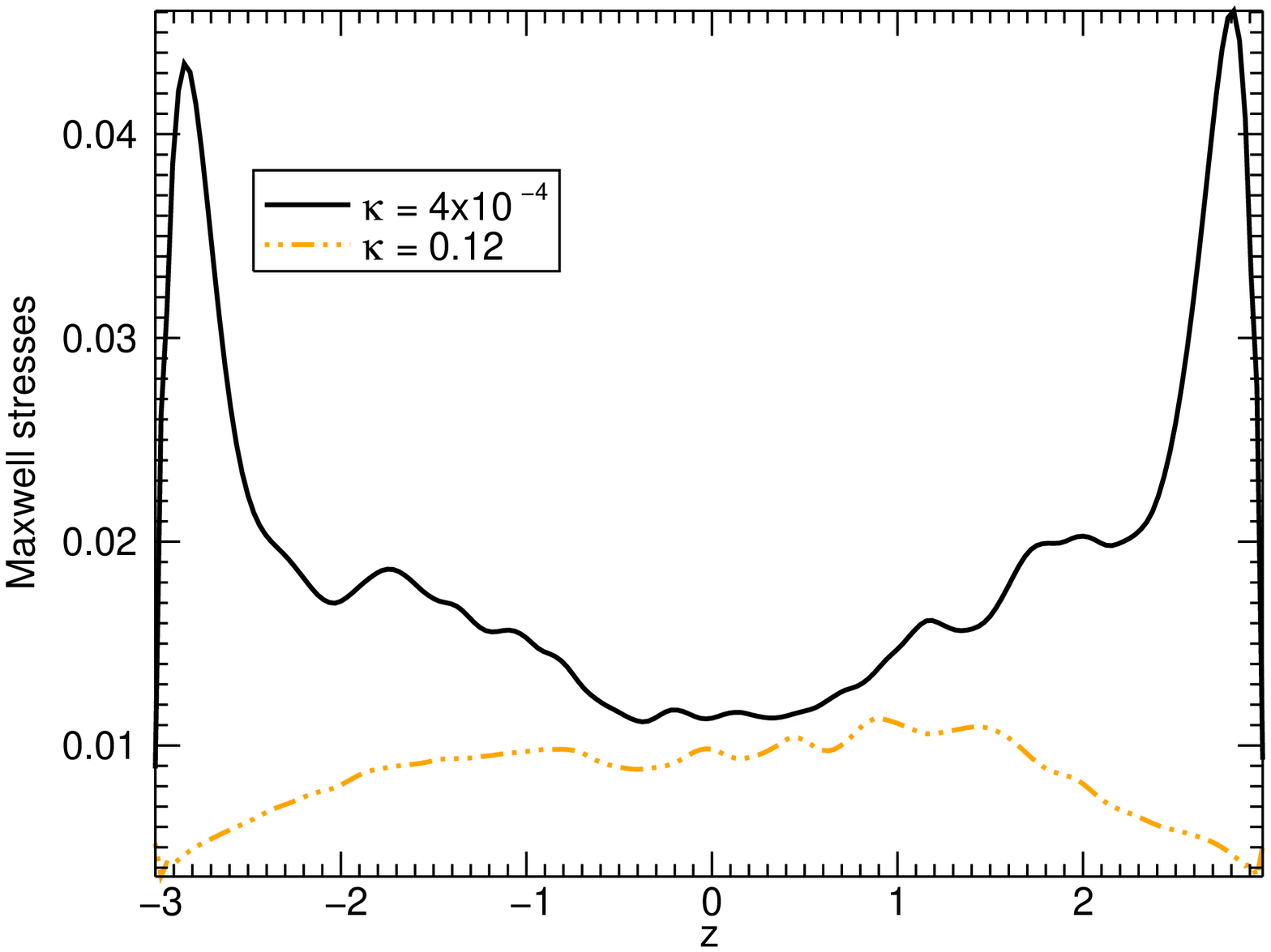} 
   \caption{Plot of the average Maxwell stresses as a function of $z$ for two cases with different t values of $\kappa$. One case (solid line) is in the convective regime, the other (dashed-dotted line) is in the conductive regime.  }
   \label{fig:av_maxw}
\end{figure}

\begin{figure}[htbp]
   \centering
   \includegraphics[width=15cm]{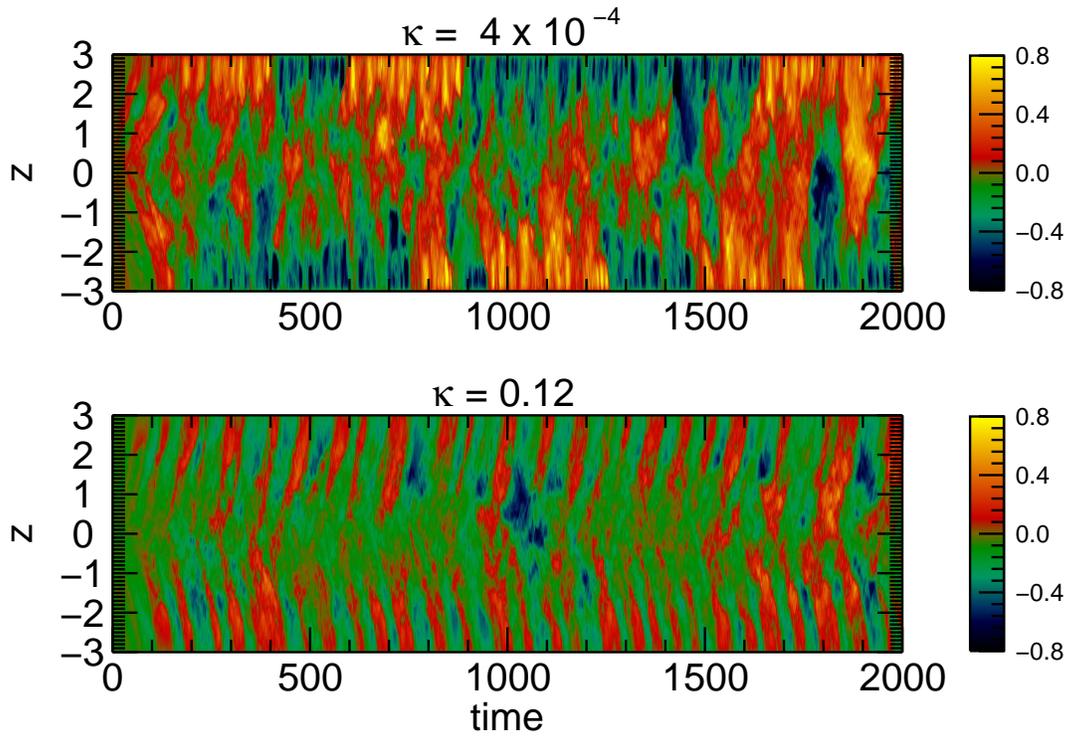} 
   \caption{Space-time diagrams of average azimuthal field. The horizontally averaged value of $B_y$ is plotted as a function of $z$ and $t$.
The upper panel corresponds to a case in the convective regime; the lower, to one in the conductive regime. The corresponding values of $\kappa$ are as indicated.  }
   \label{fig:byavg_z_t}
\end{figure}

Further evidence for two distinct types of dynamo actions operating in the two regimes can be obtained by inspection of Fig.   \ref{fig:av_maxw}. This shows the horizontally and time averaged Maxwell stresses as a function of $z$ for two cases with different values of $\kappa$ corresponding to the convective and conductive regimes. The curve for the conductive case follows the general trend of the more familiar isothermal calculations. The transport is largest in the denser central regions, steadily decreasing 
at higher values of z. This is in sharp contrast with the convective case in which the the stresses actually rapidly increase with distance from the mid-plane reaching a sharp maximum near the boundaries. In both cases, the corresponding Reynolds stresses are small and decrease steadily away from the mid-plane. The spatio-temporal behavior of the dynamo is also remarkably different in the two regimes as illustrated in Fig. \ref{fig:byavg_z_t}. The two panels show  space-time diagrams of the horizontally averaged azimuthal magnetic field as a function of $z$ and time. The lower panel, corresponding to a conductive case, displays the characteristic patterns typical of the isothermal cases  signaling the presence of cyclic activity with  magnetic structures propagating from the mid-plane to the boundaries. In the upper panel there is no evidence for cyclic activity or pattern propagation. The magnetic structures form and vanish seemingly at random with no apparent characteristic time between field reversals. Furthermore there are events in which coherent structures form that extend over the entire layer. Interestingly, for earlier times, when the layer is still close to isothermal there is some evidence for pattern propagation. From  these last two  figures it is clear that both the transport efficiency and the amount of generated toroidal flux is much higher in the convective regime than in the conductive one. 

\begin{figure}[htbp]
   \centering
   \includegraphics[width=10cm]{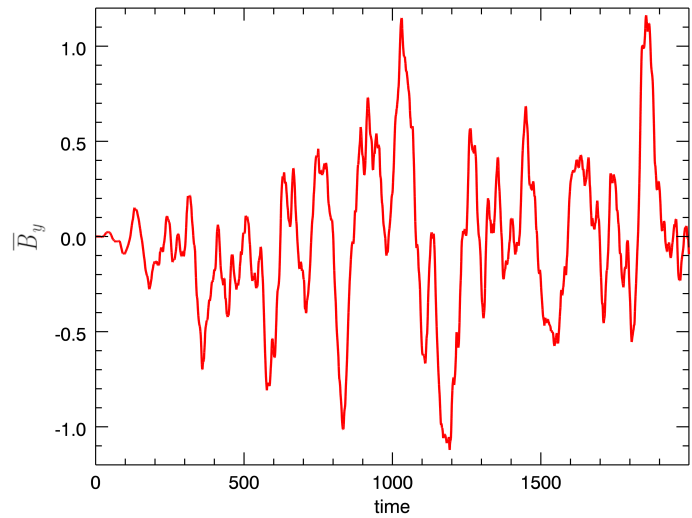} 
   \caption{Time history of $\overline B_y$ (the volume averaged $B_y$) in units of the rms value of the fluctuations. For this case $\kappa = 4 \times 10^{-3}$.}
   \label{fig:avby}
\end{figure}

A possible reason for this difference might be related to the influence of magnetic boundary conditions. It is well known that in unstratified shearing boxes the boundary conditions make a big difference to the operation of the dynamo. Periodic boundary conditions, as was mentioned in the introduction, lead to small-scale dynamo action and to the convergence problem. On the other hand, ``vertical" boundary conditions, like the ones imposed here, lead to a much more efficient dynamo that appears to scale with the system size rather than with the dissipation scale \citep{Kapyla10}. By contrast, the solutions in isothermal, stratified shearing boxes are more insensitive to the boundary conditions \citep{Davis10, Shi10, Oishi11}. This most likely is because the boundaries are located in very tenuous regions characterized by low density and high Alfv\'en speed. In the convective cases described here, the density is nearly constant as a function of height making the layer appear more ``unstratified". 
Partial support for this argument can be provided by looking at what type of dynamo is operating in the convective regime. Fig. \ref{fig:avby} shows the time history of the volume averaged value of $B_y$ (the azimuthal component) scaled in terms of the rms value of the fluctuations. Two things are worthy of notice: the average field changes sign, and its magnitude is comparable with--and actually it occasionally exceeds--that of the fluctuations. This is strongly suggestive of the operation of a  system-scale dynamo \citep{Tobias11} and should be contrasted with the corresponding isothermal case in which there is a different behavior depending on height and the ratio between average and fluctuations raises from about ten percent in the central region, where most of the transport takes place, to more substantial values in the upper and lower regions where the transport strongly declines.

\section{Conclusions} \label{conclusions}
Our main objective has been to study numerically the effects of dissipative heating and finite heat transport in determining the thermal structure of the layer and the efficiency of angular momentum transport in stratified shearing boxes with zero magnetic flux. In particular we wanted to compare with the more commonly studied isothermal case. To this end we have considered the simple case of a fluid obeying the perfect gas law and with finite (constant) thermal diffusivity. 

Our main result is to identify two distinct regimes: conductive and convective, corresponding respectively to  large and small values of the thermal diffusivity. In the conductive regime, the heat generated by  dissipation is transported through the bulk of the layer by thermal conduction, and the temperature and density have close to parabolic profiles. This appears to be in agreement with the conclusions of the recent work by \citet{Uzdensky12}. The convective regime is dramatically different. In these cases the heat is transported almost entirely by overturning motions driven by Rayleigh-Taylor type instabilities. The density profile becomes flat, and the temperature develops a  ``tent" profile with thin boundary layers at the upper and lower boundaries. There is evidence that the ``tent" profile for the temperature and the flat profile for the density are universal, in the sense that they depend solely on the properties of the turbulence and not on the values of the collisional processes.  
This last property in particular, may have important consequences for the dynamo processes. It appears that the dynamo can operate more efficiently in a layer with nearly constant density than in a corresponding layer with the same total mass and a Gaussian profile (isothermal case). This being the case,  there is a interesting feedback effect. The dynamo drives the MRI turbulence that heats the layer causing it to become Rayleigh-Taylor unstable, the overturning motions associated with the Rayleigh-Taylor instability homogenize the density allowing a more efficient operation of the dynamo, and so on until the layer settles to a universal, convective, self-regulated state. At the moment it is not clear whether the Rayleigh-Taylor driven motions contribute directly to a more efficient working of the dynamo, or they contribute indirectly by maintaining the more beneficial constant density state. Some of the similarities between the dynamo properties observed here and those in the  work of \citet{Kapyla10} in which stratification is absent suggest that it may be the constant density feature that is important. 

Finally, we remark on the natural extensions of the present model. There are two avenues that immediately come to  mind. One is to include a more realistic treatment of the thermal transport. The obvious next step is to consider thermal diffusivities that have power law dependencies on density and temperature. We anticipate that this may have some impact on the stratification in the conductive regime, but hardly any in the convective regime in which all the thermal transport is mediated by flows anyway. The other is to consider more realistic boundary conditions, like for instance those appropriate to black body radiation. Preliminary results in this direction  show that in the convective regime this choice leads to a change in the overall value of the temperature  but not in its profile. Also, the constant density profile remains unchanged. These results however are preliminary and a more thorough study is needed. Also, the assumption of impenetrable stress-free boundary conditions should be replaced by more realistic conditions in which there is a thin transition layer across which the opacity changes dramatically and the fluid goes from being optically thick to optically thin. The problem is similar to that of matching a photosphere on top of a stellar convection zone. Numerically this is extremely challenging and will be considered in future works.
However all these  extensions are secondary to the issue of convergence. In a sense, if the dynamo ceases to operate efficiently at high magnetic Reynolds numbers all bets are off. There is some room for cautious optimism since the evidence so far, is that the dynamo operating here in the convective regime is more likely to be of the system-scale type than of the small-scale type. Preliminary calculations with twice the resolutions indeed support this conjecture. However, in the end only a (very costly) convergence study will settle the issue.

\section{Acknowledgment}
This work was supported in part  by the National Science Foundation 
sponsored Center for Magnetic Self Organization at the University of Chicago.
GB, AM and PR ackowledge support by INAF through an INAF-PRIN grant. 
We acknowledge that the results in this paper have been achieved using the PRACE Research Infrastructure resource JUGENE based in Germany at  the J\"ulich Supercomputing Center.

\end{document}